# Zonal jets in equilibrating baroclinic instability on the polar beta-plane: experiments with altimetry

A. M. Matulka and Y. D. Afanasyev[a]

[a]*Memorial University of Newfoundland, St. John's, Canada*

[a]Corresponding author address:

Yakov Afanasyev

Memorial University of Newfoundland, St. John's, NL, Canada

E-mail: afanai@mun.ca

http://www.physics.mun.ca/~yakov

**Key points**

- Lab experiments reproduce eddy-driven zonal jets in baroclinic turbulence

- Scales of mesoscale menders and jets are measured

- Submesoscale filaments and eddies are observed

**Abstract**

Results from the laboratory experiments on the evolution of baroclinically unstable flows generated in a rotating tank with topographic $\beta$-effect are presented. We study zonal jets of alternating direction which occur in these flows. The primary system we model includes lighter fluid in the South and heavier fluid in the North with resulting slow meridional circulation and fast mean zonal motion. In a two-layer system the velocity shear between the layers results in baroclinic instability which equilibrates with time and, due to interaction with $\beta$-effect generates zonal jets. This system is archetypal for various geophysical systems including the general circulation and jet streams in the Earth's atmosphere, the Antarctic Circumpolar Current or the areas in the vicinity of western boundary currents where baroclinic instability and multiple zonal jets are observed. The gradient of the surface elevation and the thickness of the upper layer are measured in the experiments using the Altimetric Imaging Velocimetry and the Optical Thickness Velocimetry techniques respectively. Barotropic and baroclinic velocity fields are then derived from the measured quantities. The results demonstrate that the zonal jets are driven by "eddy forcing" due to continuously created baroclinic perturbations. The flow is baroclinic to a significant degree and the jets are "surface intensified". The meridional wavelength of the jets varies linearly with the baroclinic radius of deformation and is also in a good agreement with a modified Rhines scale. This suggests a linear dependence of the perturbation velocity in the equilibrated baroclinically unstable flow on the β-parameter.

## 1. INTRODUCTION

This study is motivated by observations of zonal jets in the oceans [Maximenko et al., 2005; 2008]. These jets were revealed by time-averaging of the satellite altimetry data; they are typically obscured by a stronger signal from energetic mesoscale eddies. The jets are a subject of current discussion in the literature. However, the physical mechanisms which contribute to the jets' formation, as well as the details of their dynamics, are not yet fully understood.

Several approaches were used to explain the creation of zonal jets. One of them uses a concept of β-plumes. Observational evidence indicates that the jets often originate at the eastern boundary of oceans [Centurioni et al., 2008; Ivanov et al., 2009]. β-plumes are formed via the radiation of Rossby waves by perturbations at the boundary. Quasi-steady meanders of coastal currents are often tied to coastal topography. They can radiate Rossby waves and shed eddies westward. A β-plume created by almost zero frequency Rossby waves is an essentially zonal circulation/gyre which consists of two jets flowing in opposite directions westward of a localized perturbation which generates the β-plume. The jets are reconnected around the area of the perturbation as well as in front of the plume. Recent laboratory experiments [Afanasyev et al., 2011; Slavin and Afanasyev 2012] demonstrated how the β-plume mechanism works. In these experiments, either a baroclinically unstable coastal current or an unstable patch of warm fluid were used to show that Rossby waves are initially radiated by the perturbations and form β-plumes which, in turn, provide pathways for the baroclinic eddies to migrate westward.

When the entire domain is filled with eddies, the circumstances are somewhat different. A general concept of interplay between eddies at smaller scales and Rossby waves at larger scales can be used to understand the transfer of energy towards zonal modes. This concept was offered by Rhines [1975] who found a dividing length scale between eddies and waves. This scale is also viewed as an estimate of the meridional wavelength of zonal jets. It was demonstrated that the Rhines scale is quite universal and is applicable in many circumstances, although different modifications and limitations of this scale were discussed as well. Thus, jets can be sustained by eddies via the Rhines mechanism and, therefore, must occur where eddies are plentiful and energetic. Mesoscale eddies in the oceans (as well as in the atmosphere) are created mainly by baroclinic instability. The geography of the most baroclinically unstable regions includes the Antarctic Circumpolar Current (ACC), and western boundary currents such as the Gulf Stream and Kuroshio and their extensions [Smith, 2007]. Observations show that multiple zonal jets are indeed created in these regions and are quite prominent there [Maximenko et al., 2005, 2008; Sokolov and Rintoul, 2007; Thompson, 2008].

Berloff et al. [2009 a, b] suggested yet another way which leads to creation of zonal jets. The authors studied a quasigeostrophic model of flow in a zonal channel. When background flow with vertical shear is imposed, the flow is baroclinically unstable. The primary instability mode is in the form of so-called noodles, alternating meridional jets. The noodles are, in turn, subject to a secondary instability in the form of zonal jets. Linear stability analysis can be used to describe both the primary and secondary instabilities thus providing an initial path from baroclinic instability towards zonal jets. However, the equilibrated/saturated flow with continuous baroclinic instability and zonal jets must include nonlinear dynamics in the form of eddy fluxes. The question then is whether the Rhines mechanism works in the equilibrated flow or the presence of underlying baroclinic instability makes the dynamics significantly different. The measurements of the meridional jet scaling by Berloff et al. [2009 a, b] showed that the Rhines scaling did not work universally but only within a certain range of parameters. Note that an earlier numerical study by Panetta [1993] did confirm the Rhines scaling within a similar parameter range. Recent numerical simulations of a two-layer system by Williams and Kelsall [2015] showed an agreement with the Rhines scaling modified by inclusion of the baroclinic radius of deformation.

The properties of zonal jets generated by baroclinic instability must be related to the properties of the mesoscale variability which includes baroclinic meanders, filaments and eddies. Although the linear mechanism of baroclinic instability is well understood, its nonlinear properties in a saturated state are still a subject of investigation. A recent study by Radko et al. [2014] offers a theoretical model which allows one to predict certain properties of the variability generated by baroclinic instability. The theory is based on an assumption that instability saturates when the growth rates of the primary instability (meridional jets, noodles) and the secondary instability (zonal jets) are comparable. The theory (in combination with numerical simulations) allowed the authors to evaluate the meridional eddy flux in the saturated baroclinic flow.

In this experimental study we model a system which includes lighter fluid in the South and heavier fluid in the North. This initial stratification results in a slow meridional circulation and fast mean zonal flow. We employ a two-layer system for simplicity. Velocity shear between the layers results in baroclinic instability which equilibrates and, due to interaction with β-effect generates zonal jets. This system is archetypal for various geophysical systems including the Earth's atmosphere where the general circulation and jet streams are due to temperature gradient between the equator and the poles, or the Antarctic Circumpolar Current which is created by the meridional density gradient between the cold Antarctic water and warm subtropical water. The purpose of this study is to provide an experimental evidence for the zonal jet scaling and study its relation to the Rhines scale and to the characteristics of the baroclinic instability. In what follows, we describe the phenomenology of our laboratory flows including the formation of zonal jets, "noodles" and small-scale eddies. We also measure nonlinear eddy fluxes in order to understand jet forcing in the upper and lower layers. The results of the experiments can be used as a "benchmark" for comparison with numerical simulations.

There were relatively few experimental studies where baroclinic zonal jets were investigated in detail. Read et al. [2007] studied a flow where zonal jets were generated by small-scale convection. The surface of fresh water was sprayed with salty water; the drops of salty water formed small convective plumes descending to the bottom. Small-scale turbulence generated by the convection then caused the formation of zonal jets. Similar experiments but with a thermal forcing were also described in Zhang and Afanasyev [2014]. Heating wire on the bottom of the tank was used to generate convection. However, in both cases the large-scale flows were mostly barotropic despite the fact that forcing was baroclinic in nature. The small-scale forcing presumably generates an inverse cascade which then, in accordance with the Rhines mechanism, transfers energy to Rossby waves and created jets. Smith et al. [2014] performed a series of experiments in a classic setup with a differentially heated annulus. The differential heating drives a meridional circulation and creates an approximately geostrophic mean zonal flow. The flow is unstable with respect to the baroclinic instability. The authors observed the formation of highly meandering zonal jets; the number of jets increased with the rotation rate of the table. They showed that time-averaged jets were driven by the Reynolds stress forcing. The scaling of jets agreed reasonably well with the Rhines scaling. The velocity was measured by Particle Imaging Velocimetry (PIV) in a horizontal cross section of the flow. However, neither in the experiments by Smith et al. [2014] nor in the previous experiments the baroclinic structure of the flow was measured in sufficient detail.

In this study we use a recently developed experimental technique, Altimetric Imaging Velocimetry (AIV) together with the Optical Thickness Velocimetry (OTV) to overcome the shortcomings of the earlier techniques and to measure both barotropic and baroclinic components of the flow and thus obtain a full information about the system.

In Sec. II of this paper, we describe the setup of the laboratory apparatus as well as the AIV and OTV techniques. In Sec. III the results of the laboratory experiments are reported. Concluding remarks are given in Sec. IV.

## 2. LABORATORY SETUP AND TECHNIQUE

A circular tank of radius $R = 55$ cm filled with saline water of depth $H_0 = 10 - 12$ cm was used in our experiments (Figure 1). The salinity, $S$, of water in the tank varied between the experiments in the range between $S = 5$ ppt and 65 ppt. The tank was installed on a rotating table and was rotated anticlockwise at a constant angular rate $\Omega = 2.32$ rad/s.

In total we performed 21 experiments where the flows were generated by a source of fresh water located along the wall of the tank at the surface of saline water layer. Fresh water was pumped into the tank through a perforated pipe which was attached to the wall of the tank and extended for approximately one quarter of the circumference of the wall. The water was supplied by a pump which delivered 20 liters of water in approximately 6 min. The fresh water from the source was lighter than saline water in the tank and it spread on top of the saline water forming a fresh surface layer. All measurements in our experiments were conducted after the forcing (pumping of fresh water) stopped and the flow was slowly relaxing. At that time the flow consists of two-layers, where the surface layer of fresh water extends over the entire area of the tank except, perhaps, a small area in the center.

The total depth of the fluid in a rotating tank is parabolic when in solid-body rotation due to the balance between the pressure gradient directed towards the axis of rotation and the centrifugal force. The height $h$ of the water surface varies quadratically with the distance $r$ from the axis of rotation

$$h(r) = H_o + \frac{\Omega^2}{2g}\left(r^2 - \frac{R^2}{2}\right) \quad (1)$$

where $H_o$ is the depth of the layer in the absence of rotation and $g$ is the gravitational acceleration. Quadratic variation of depth can be used to model the variation of the Coriolis parameter which occurs on a rotating planet due to its sphericity. Dynamical effects of decreasing Coriolis parameter and increasing depth are equivalent due to the conservation of potential vorticity (PV). Thus we have a so-called "topographic" β-plane. In our particular case, when the depth is given by Eq. (1), the β-plane includes the North pole and the equivalent Coriolis parameter varies quadratically with the distance from the pole in the first approximation

$$f \approx f_0 - \gamma r^2, \tag{2}$$

where the coefficient $\gamma$ is

$$\gamma = \Omega^3 / (gH_0). \tag{3}$$

This approximation is called a polar $\beta$-plane. A regular (linear) $\beta$-plane can be introduced locally, in the midlatitudes of the tank. For this purpose a local Cartesian system is defined such that its origin is at some reference "latitude" given by the distance from the pole $r_0$; the positive *x*-direction is East and positive *y*-direction is North. The equivalent Coriolis parameter then varies linearly with *y* as

$$f \approx f(r_0) + \beta r, \tag{4}$$

where the Coriolis parameter at the reference latitude is

$$f(r_0) = f_0 - \gamma r_0^2 \tag{5}$$

and

$$\beta = 2r_0\gamma. \tag{6}$$

We employ two different optical techniques to measure the characteristics of the flows. The first technique is called the Altimetric Imaging Velocimetry (AIV). It allows us to measure the slope of the surface elevation. The second technique, Optical Thickness Velocimetry (OTV) measures the thickness of one of the layers. The slope of the surface elevation and the thickness of the layer are then used to obtain barotropic and baroclinic velocities respectively. The hardware for these techniques is shown schematically in Figure 1 and includes a video camera, a light source with a color mask located above the tank and a light box under the tank. The video camera records up to 10 frames per second of size $2300 \times 2300$ pixels. These characteristics specify the spatial and temporal resolution of our measurements. In what follows, we briefly describe both of the techniques and refer to Afanasyev et al. [2009] for further details.

AIV uses the parabolic surface of water as a mirror. The camera observes the reflection of the multicolor mask; if the surface is perturbed by the pressure perturbations due to the flow, the reflected color allows us to determine the slope. A typical altimetric image of the flow is shown in Figure 1 b. We measure two components of the gradient $\nabla \eta = (\partial \eta / \partial x, \partial \eta / \partial y)$ of the surface elevation $\eta$ in the horizontal plane $(x, y)$. The surface elevation is defined with respect to the parabolic surface given by Eq. (1). The velocity of the flow can then be obtained in a geostrophic approximation

$$\mathbf{V}_g = \frac{g}{f_0} (\mathbf{n} \times \nabla \eta), \tag{7}$$

or in a quasi-geostrophic (QG) approximation

$$\mathbf{V} = \mathbf{V}_g - \frac{g}{f_o^2} \frac{\partial}{\partial t} \nabla \eta - \frac{g^2}{f_o^3} J(\eta, \nabla \eta), \tag{8}$$

where $\mathbf{n}$ is the vertical unit vector and $\mathbf{V}$ is the horizontal velocity vector. While the measured surface slope is precise (within the experimental error), the "textbook" criteria of validity of the geostrophic or QG approximations applies for velocity. The velocity, given by Eqs. (1) or (2) is, in fact, a barotropic velocity of the flow.

In order to obtain the baroclinic velocity, we employed the OTV technique. OTV is based on the absorption of light by a dye dissolved in water. The upper layer which contained the water supplied by the source was dyed by a red (or green) food dye in our experiments. Uniform white light provided by the light box under the tank passed through the dyed layer and was observed by the camera (Figure 1 c). The color of the layer varied with its thickness. Thus, the thickness of the dyed upper layer, $h_1$, can be measured by the intensity of its color. Baroclinic geostrophic velocity due to the variation of the thickness of the layer can be easily obtained as

$$\mathbf{V}_{gbc} = -\frac{g'}{f_o}\left(\mathbf{n} \times \nabla h_1\right), \qquad (9)$$

where $g' = g(\rho_2 - \rho_1)/\rho_1$ is the reduced gravity, $\rho_1$ and $\rho_2$ are the densities of the upper and lower layer respectively. QG baroclinic velocity, $\mathbf{V}_{bc}$, can be obtained from Eq. (8) by substituting $\eta$ by $-h_1$.

AIV and OTV can be used together if we alternate the light under the tank and the light with color mask above the tank. The AIV and OTV images are separated by a short time interval (determined by the frame rate of the camera) such that the measured barotropic and baroclinic velocity fields are almost simultaneous. Thus, the velocity in the upper layer is equal to the barotropic velocity, $\mathbf{v}_1 = \mathbf{V}$, while the velocity in the lower layer is given by the sum of the barotropic and baroclinic velocities, $\mathbf{v}_2 = \mathbf{V} + \mathbf{V}_{bc}$.

## 3. RESULTS

We performed twenty one experiments with different salinity where the baroclinic radius of deformation, defined as $R_d = (g'H_0/2)^{1/2}/f_0$, varied between 0.66 cm and 2.4 cm. The scale of meanders and jets observed in our experiments varied accordingly. However, a typical scenario of the flow evolution was similar in all experiments. In what follows we describe the general evolution of the flow, discuss the eddy forcing of the zonal jets and present the experimental data on the scaling of the jets and baroclinic meanders.

### 3.1 General circulation

The source of fresh water at the wall of the tank creates a strong coastal current which flows counterclockwise such that the wall is on its right (Figure 2). The current is unstable with respect to a baroclinic/frontal instability and forms meanders. The current encircles the tank; the fresh water supplied by the source, gradually fills the interior of the tank. As a result, a fresh water layer forms at the surface, on top of the saline layer. By the time when the forcing is stopped, the fresh water covers the entire area of the tank. After that, the flow continues to evolve, albeit more slowly. Mean circulation in both the top and bottom layers is easy to understand. The surface is slightly elevated near the wall. As a result, the coastal current in the upper layer flows in the cyclonic direction such that a simple balance between the Coriolis force directed towards the wall and the pressure gradient acting in the opposite direction, is maintained. The upper layer is thicker at the wall of the tank than in its center (it adjusts towards a layer of uniform thickness on a very long time scale). The pressure in the lower layer is relatively low near the wall because of the relatively thick layer of (lighter) fresh water on top and relatively high in the center of the tank. Simple geostrophy then suggests that an anticyclonic gyre must be created in the lower layer. Thus, the top and bottom layers rotate in the opposite directions. The velocity shear is created and maintained by the baroclinic structure of the two-layer system. Theory predicts that a velocity shear in combination with the density difference between the layers should result in a baroclinic instability. Indeed, the instability was observed in all of our experiments; it persisted for a long time after the forcing stopped.

Baroclinic instability continuously creates meanders and filaments which protrude in the meridional (radial) direction. Figure 2 shows that long narrow filaments are a dominant feature of the flow. It is worth noting here that eddies are quite rare in these flows. Relatively large eddies/gyres form mainly by the meanders of the coastal current in space between the current and the wall of the tank. Eddies are less often observed in the interior of the domain. Relatively small eddies form spontaneously within filaments (Figure 2 e, f) but their lifetime is quite short (these eddies will be discussed in more detail later). The meridional motion of the filaments is affected by the β-effect. As a result, zonal jets are formed. The zonal jets of alternating direction can be clearly seen both in the snapshots of the flow (Figure 2c, d) or in the Hovmoeller diagrams as shown in Figure 3. Distributions of the azimuthal velocity along the radial direction (at the azimuthal angle $\theta = 180^0$) were recorded at different times in two experiments. The experiments were with different salinity of the lower layer, $S = 30$ ppt and $S = 10$ ppt respectively. In both experiments the coastal jet was quite prominent; in Figure 3 a and b the jet is a band of dark red color near the wall, at radius $r \approx 50$ cm. Several jets flowing eastward (red/yellow) and westward (blue/cyan) can also be identified in the interior of the tank. The comparison of panels a and b shows that the flow in the higher salinity (higher $R_{bc}$) experiment is stronger and the scales of meanders and jets are larger, as one can expect. Although jets are disturbed by perturbations (especially in the higher salinity experiment), they are persistent and remain approximately in the same location for a long time. Diagrams in panels c and d in Figure 3 show the radial velocity recorded along the azimuthal direction, a circle of radius $r = 2R/3 \approx 36$ cm. The circle is well away from the coastal jet such that the dynamics of the flow there is not affected by the wall. The diagrams reveal some interesting details of the flow. Firstly, a periodicity of the perturbations along the circle is quite clear. These perturbations are the meanders/filaments occurring due to baroclinic instability. By counting the maxima or minima of the radial velocity the wavelength of the perturbations can be measured (and will be discussed later). Secondly, the slope of colored bands indicates the direction and the speed of phase propagation in the azimuthal direction. In the beginning, in the first 100 s, baroclinic meanders rapidly propagate to the west (towards smaller $\theta$) while later the direction of their propagation changes to eastward. This indicates that the system goes through an adjustment after the pumping is stopped. After this relatively short adjustment, the evolution of the system is much slower.

Figure 4 shows the root-mean-square (rms) barotropic and baroclinic velocities as functions of time in three experiments with different salinity. The velocities were measured within a ring including the "midlatitudes" of the domain in order to avoid the coastal current at the "equator" and the area around the North pole. The rms velocity plots also indicate the period of the adjustment in the beginning and the period of the very slow evolution later in the experiments. In fact, both barotropic and baroclinic velocities remain almost constant during the latter phase. This feature requires some explanation. Dissipation in the rotating fluid can be interpreted in terms of a bottom Ekman friction and "regular" friction due to viscosity in the bulk of the fluid. Contributions due to each effect can be measured for a particular flow [e.g. Afanasyev and Craig, 2013; Zhang and Afanasyev, 2014]. Typically the relative contribution of the Ekman friction increases at later phase when turbulence decays to a certain degree. The energy of the flow then decays exponentially with time such that the energy decay rate is constant. The fact that in our present experiments the rms velocity and hence the energy of the flow remain constant indicates that the baroclinic potential energy is released and converted to the kinetic energy at the same rate as it dissipates. Thus, we have an approximately stationary system which assures that our measurements will have a more universal significance rather than being just a snapshot of a time dependent flow. In what follows we describe the flow characteristics that were measured only during the stationary phase in each experiment. Plots in Figure 4 also show that velocities are higher in the experiments with high salinity which can be easily understood because the available potential energy is higher in these flows. The ratio of the baroclinic to barotropic rms velocity (Figure 4 c) can serve as a characteristic of the baroclinicity of the flow. This ratio is typically greater than unity except for the experiments with lowest salinity, $S < 15$ ppt, that indicates that the flows are baroclinic to a significant degree. The ratio increases with salinity as one might expect. It also remains approximately constant during the stationary phase in most of the experiments that indicates that the flows do not become more barotropic with time.

Figure 5 shows typical snapshots of the barotropic azimuthal and radial velocity in two experiments with different salinity. The velocity was measured during the stationary phase in both experiments. Well-formed zonal jets can be clearly identified in the images of the azimuthal velocity. The difference between the meridional scales of the jets is apparent; in the experiment with larger salinity difference (larger $R_d$) the jets are wider than those in the experiment with relatively small salinity. The jets meander; the zonal wavelength of the meanders also exhibits the dependence on the radius of deformation. We discuss this dependence in one of the following sections. Here, however, it is worth noting that these meanders are quite coherent in the meridional direction. Panels c and d in Figure 5 show the radial velocity in polar coordinates ($r$, $\theta$) with the origin at the center of the tank. The velocity field exhibits a pattern of alternating jets flowing in the radial (meridional) direction. These meridional jets are not obvious in the vector plots of velocity (e.g. Figure 2 c) because of the dominance of the zonal motion but they are a persistent feature of the flow as long as the baroclinic instability is active. It is also worth noting that the mean barotropic flow in most of the experiments has a dipolar component. Azimuthal velocity is stronger at approximately 4 o'clock in Figure 5 a, b and weaker on the opposite side of the tank. The radial velocity has maximum at approximately 250 degrees or 7 o'clock. This defines the axis of the dipole with a cross-polar flow directed from 1 to 7 o'clock. The exact reason for the appearance of this circulation is unclear. It is possible that it occurs due to initial forcing conditions and then persists during the experiment. However, this secondary circulation does not appear to have a significant effect on the properties of jets apart from modulating their magnitude.

**3.2 Eddy forcing**

Zonal jets were observed to be a robust feature in our experiments. Herein, we analyze the forcing of the jets by meanders, filaments and eddies created in the baroclinically unstable flow. This forcing is usually called "eddy forcing"; it is described statistically by presenting the flow in terms of its mean characteristics and perturbations without regard to a particular character of the perturbations. In a two-layer system, the mean zonal momentum balance in the statistically steady state is given by [e.g. McWilliams, 2006]

$$\frac{\partial \bar{u}_n}{\partial t} \approx \begin{cases} -\dfrac{D_{1.5}}{\bar{h}_1} - \dfrac{\partial R_1}{\partial y}, & n=1 \\ \dfrac{D_{1.5}}{\bar{h}_2} - \dfrac{\partial R_2}{\partial y} - f_{bot}, & n=2 \end{cases}, \qquad (10)$$

where overbar denotes time mean values, $\bar{h}_1$ and $\bar{h}_2$ are the thicknesses of the upper and lower layer respectively and eddy fluxes are defined as

$$D_{1.5} = -f_0 \overline{v'_{1.5} \eta'_{1.5}}, \qquad (11)$$

$$R_n = \overline{u'_n v'_n}. \qquad (12)$$

Index 1.5 refers to the quantities at the interface between the layers; $v_{1.5} = (v_1 + v_2)/2$ and $\eta'_{1.5} = \bar{h}_1 - h_1$ is the elevation of the interface. $D_{1.5}$ is the form drag which occurs because of the undulated interface. The form drag describes the exchange of momentum between the layers; when one layer is accelerated by this term, another layer is decelerated. $R_n$ is the Reynolds stress and the forcing is given by its horizontal divergence. The term $f_{bot}$ in Eq. (10) represents the bottom Ekman friction. Friction due to regular viscosity is not included in Eq. (10) but is certainly a factor in both layers. In a steady state there is no acceleration, $\partial \bar{u}_n / \partial t = 0$, and the right-hand side of Eq. (10) must be balanced in both layers.

Let us consider the eddy forcing components using one particular experiment with $S = 45$ ppt as an example. Figure 6 shows the fields of barotropic and baroclinic azimuthal velocity. Apart from the coastal jet, at least two prominent eastward jets can be easily identified in the snapshot of the barotropic (upper layer) velocity field (Figure 6 a). Baroclinic jets approximately mimic the barotropic ones but they flow in the opposite direction. Azimuthal component of the Reynolds stress is given in Figure 6 c. The eddy forcing is concentrated along the jets and the forcing in the coastal jets is much weaker than that in the interior jets which indicates that their dynamics is different. The periodicity of stress in the zonal direction reflects the pattern of the baroclinic meanders that create the stress. Similar pattern was observed for the form drag $D_{1.5}$ (not shown here). The radial profiles of azimuthal velocities and eddy forcing components in the upper and lower layers are shown in Figure 7. The time averaging was performed over the interval of 50 s to calculate the eddy forcing terms. The averaging in the zonal direction over a sector of approximately $90^0$ was then used to obtain the profiles. Barotropic velocity profile (red in Figure 7 a) shows a strong jet at $r \approx 30$ cm as well as weaker jets in the interior and a strong coastal jet near the wall. The baroclinic velocity (green) is almost a mirror reflection of the barotropic one. The profile of the lower layer velocity (blue) which is given by the sum of the barotropic and baroclinic components indicates, however, that there are no prominent jets in the lower layer. The jets are weaker there and somewhat narrower in the meridional direction. They are imbedded in the anticyclonic mean circulation which is approximately solid-body type in the interior of the domain. Panels b and c in Figure 7 show a balance of the eddy forcing components given by the terms in the RHS of Eq. (10) in both layers. In the upper layer, the Reynolds stress (red) is the main driving component; it correlates very well with the azimuthal velocity. The Reynolds stress is approximately balanced by the form drag. The form drag is negatively correlated with the Reynolds stress but is somewhat smaller in magnitude. The residual of the sum of these two stresses is most likely balanced by viscosity. The dynamics in the lower layer is less clear. The form drag is less significant there because of the lower layer is relatively thick. The magnitude if the Reynolds stress is quite high but the correlation between the stress and the lower layer velocity is not very strong. Typical radial (meridional) scale of the stress is, however, similar to that of the velocity. A significant part of the balance in the lower layer should also be the Ekman friction (apart from regular friction) which was not accounted for here.

Figure 7 d shows a typical time-mean profile of the potential vorticity in the upper layer, defined as $q_1 = (f_0+\zeta)/h_1$, where $\zeta$ is the relative vorticity in the upper layer. We measured PV in order to see how well it is mixed by the flow. One of the existing theoretical concepts of zonal jets predicts a formation of the PV "staircase" assuming that jets provide strong mixing of the PV or other material properties. In our experiments, we found that jets do provide strong mixing where regions of uniform PV or even inverted PV can be formed. However, strong gradients of PV between the uniform areas fail to form, PV varies quite gradually there. This property of the flow is in accord with the results of numerical simulations by Berloff et al. [2009 a] where the authors observed "washboards" instead of "staircases" of PV.

**3.3 Scaling of zonal jets and baroclinic meanders**

In this section we summarize the measurements from all the experiments to show how the meridional wavelength of jets and zonal wavelength of baroclinic meanders very with control parameters of the flow (Figure 8). The zonal wavelength, $\lambda_{zon}$, of the finite-amplitude meanders created by baroclinic instability was measured when the flow was (quasi-)steady in each experiment. The wavelength was measured along a circle of radius $r = 2R/3$ by counting the bands in the space-time diagrams of radial velocity such as those in Figure 3 c, d. The results demonstrate that the variation of $\lambda_{zon}$ with the baroclinic radius of deformation, $R_d$, is well described by a linear function, $\lambda_{zon} = 12R_d$ (black symbols and line in Figure 8 a). Typically the wavelength of the most unstable mode of baroclinic instability is approximately 4 - $6R_d$, but the value of $12R_d$ obtained here is not particularly surprising for finite-amplitude perturbations.

The meridional wavelength, $\lambda_{jet}$, of zonal jets was also measured in a steady regime in all experiments. We used space-time diagrams of zonal velocity (e.g. Figure 3 a, b) for this purpose. The wavelength was only measured in mid-latitudes of the domain in order to exclude the coastal jet and the polar region from the measurements. Similar to baroclinic meanders, jet scaling also demonstrates a linear dependence on the radius of deformation, $\lambda_{jet} = 4.4R_d$ (blue symbols and line in Figure 8 a). Since the jets are driven by the eddy forcing which is due to baroclinic meanders, the dependence of the jet wavelength on $R_d$ is not unexpected. However, there is another control parameter in the flow that should determine the jet scale. This parameter is due to the β-effect and is given by

$$L_{Rh} = 2\pi(V_{rms}/\beta)^{1/2}. \qquad (13)$$

$L_{Rh}$ is the well-known Rhines scale and was defined as a dividing length scale between an isotropic turbulence (eddies) and linear Rossby waves [Rhines, 1975]. This transition scale has been viewed as an estimate of the meridional scale of the jets in different flows [e.g. Vallis and Maltrud, 1993; Huang et al., 2001; Danilov and Gurarie, 2004]. It was also shown to apply to laboratory flows where barotropic [Afanasyev and Wells, 2005; Zhang and Afanasyev, 2014] or baroclinic [Read et al., 2007; Slavin and Afanasyev, 2012] jets were observed. Here we test the applicability of the Rhines scale as well. Taking into account that Rhines' original theory did not include a mean flow, we use the rms radial velocity, $v_{rms}$, instead of the total velocity to exclude the mean zonal flow. In our circumstances $v_{rms}$ provides a suitable measure of the translational velocity of baroclinic meanders and filaments which drive the zonal jets. Figure 8 b shows that the wavelength of the jets varies linearly with $L_{Rh}$, $\lambda_{jet} = 0.51\ L_{Rh}$.

From the first glance the fact that $\lambda_{jet}$ is determined equally well by two control parameters might seem controversial. Indeed, one of the parameters includes β-effect (which, of course, is necessary to create jets) and the other does not. However, this simply means that certain characteristics of the baroclinically unstable flow in its saturated state are affected by the β-effect. The Rhines scale includes $v_{rms}$ which is not an external control parameter but rather determined by the dynamics of the flow. Thus, one can expect the dependence of the form

$$v_{rms} \sim \beta R_d^2. \tag{14}$$

Figure 8 c shows that the dependence between $v_{rms}$ and $\beta R_d^2$ is approximately linear (although scatter is significant) with a coefficient of proportionality 1.5.

## 3.4 Eddies

Although jets, meanders and filaments are the most prominent features of the flow in our experiments, eddies were observed as well. These eddies are cyclonic they are of relatively small scale. Eddies are created and then disappear spontaneously. They are more likely to appear and live longer in the flows with higher salinity difference between the layers. Their lifetime is only a few periods of rotation of the table ($T$ = 2.7 s) or a few "days" if compared with a rotating planet, in the flows with relatively low salinity, $S = 10 - 15$ ppt, but increases to one "week" or more when $S = 65$ ppt. The scatter in the size of eddies was significant and we were not able to observe any clear dependence on the radius of deformation. These small-scale eddies are created mostly in the beginning of the experiments when the flow is stronger and they are practically nonexistent at the later phase.

Figure 9 shows a flow with relatively high salinity, $S = 45$ ppt. Several eddies can be observed in the snapshot of potential vorticity in the upper (Figure 9 a) and lower (Figure 9 a) layers. The eddies are formed by filaments which are bend due to interaction with other elements of the turbulent flow. Breaking of Rossby waves or shear instability are likely mechanisms of their formation. In videos of potential vorticity (not shown here) one can see clearly a polar vortex (not unlike that in the Earth's atmosphere) with filaments extruding from it. The filaments fold in the waves and create eddies. Most often these eddies are almost immediately absorbed again by the polar vortex but sometimes they are ejected and can survive if they get into a region of a relatively calm flow. One of such relatively long-lived eddies are indicated by an arrow in Figure 9 a. If we zoom-in into the eddy we can observe its characteristics in more detail. Barotropic and baroclinic velocity fields in the eddy are shown in Figures 9 c and d respectively. Their radial profiles obtained by averaging over the azimuthal angle in polar coordinates centered in the vortex are given in Figure 10. Barotropic and baroclinic velocities are of opposite sense; baroclinic component almost compensates the barotropic one such that the motion in the lower layer is relatively weak. Thus the eddy is "surface intensified" and strongly baroclinic. The radius of maximum velocity is quite small, $r_{max} \approx 1$ cm. The profile of surface elevation, $\eta$ (Figure 10 b) shows that the water surface is depressed above the eddy. The thickness of the upper layer, $h_1$ (Figure 10 c) is smaller in the center of the vortex than in its periphery. In fact, the thickness image in Figure 9 d shows that there is no green-dyed water in the center of the vortex; thus there is a "hole" in the upper layer. This means that there is a circular front separating saline water at the periphery and fresh water in the center of the vortex. The amplitude of the thickness variation is approximately $\Delta h_1 \approx 0.4$ cm, while the amplitude of the surface elevation is $\Delta \eta \approx 0.025$ cm. Taking into account that the reduced gravity in this experiment is $g' = 34$ cm/s$^2$, the observed amplitudes explain the fact that the flow in the lower layer is almost compensated. The eddy does not have a solid-body-like rotating core; the relative vorticity, $\zeta$, is not constant near the center of the eddy but decreases continuously (black line in Figure 10 a). The ratio $\zeta/f_0$ is greater than unity in the center which indicates that the eddy is strongly ageostrophic.

## 4. CONCLUSIONS

In our experiments we observed the formation of zonal jets in a baroclinically unstable two-layer flow on the (topographic) polar β-plane. The mean flow in this system is created due to the relative abundance of lighter fluid in the South (at the wall of the tank). The mean flow is cyclonic in the upper layer and anticyclonic in the lower layer. The flow is unstable; baroclinic instability continuously creates meanders and filaments in the entire domain.

The analysis of eddy fluxes showed that "midlatitude" zonal jets in the upper layer are driven by the Reynolds stresses which are provided by the baroclinic meanders. The jets in the lower layer are partially compensated and weaker than those in the upper layer. The momentum exchange between the layers is provided by the form drag. The flow remains considerably baroclinic throughout the experiment.

Measurements show that radial velocity perturbations are quite coherent along the radial direction; they form a regular pattern of alternating meridional (radial) jets. These observations are in accord with the theoretical/numerical results where primary baroclinic instability is in the form of so-called noodles [Berloff et al., 2009 b].

Although baroclinic meanders occasionally form closed or partially closed circulations, "true" eddies that separate and exist on their own for some time were quite rare in our flows. Eddies that we observed, were only formed in the initial phase of the experiments when the flow was quite strong and density fronts were formed. These eddies were exclusively cyclonic, small-scale; they were, in fact, "holes" in the upper layer. Their lifetime was in the range from a few days to more than a week depending on the salinity difference between the layers.

The meridional wavelength of the eddy-driven zonal jets varies linearly with the baroclinic radius of deformation. It was also found that the jets wavelength is in good agreement with the Rhines scale defined with the rms radial velocity. This suggests a dependence of the rms radial velocity in the equilibrated baroclinically unstable flow on the parameter $\beta R_d$.

Our observations show that the flows (especially in the early stages of their evolution) consist of long and narrow filaments with fast currents flowing along them. If baroclinic meanders of scale of approximately $12R_d$ in our experiments can be considered as an analogue of oceanic mesoscale eddies (typically 100 – 300 km in horizontal extent), the smaller scale filaments and cyclonic eddies in the laboratory flows can be qualified as submesoscale motions. Submesoscale dynamics in the ocean (features of size 1 – 100 km) has been a subject of a number of recent numerical [e.g. Qiu et al., 2014; Sasaki et al., 2014] and field [e.g. Callies et al., 2015] studies. These studies showed that submesoscale eddy field is quite energetic and can significant impact the circulation of larger scale. The submesoscale features can provide strong vertical velocities and thus play an important part in the vertical exchange in the ocean. Laboratory experiments provide an alternative way to study submesoscale motions. We hope to address the mechanisms of their formation and the spectral characteristics of the submesoscale motions in a subsequent paper.

**ACKNOWLEDGMENTS**

YDA gratefully acknowledges the support by the Natural Sciences and Engineering Research Council of Canada. Experimental data is available on request from YDA (afanai@mun.ca).

**References**

Afanasyev, Y. D., S. O'Leary, P. B. Rhines, and E. G. Lindahl (2011), On the origin of jets in the ocean, *Geoph. Astroph. Fluid Dyn.*, doi: 10.1080/03091929.2011.562896.

Afanasyev, Y. D., P. B. Rhines, and E. G. Lindahl (2009), Velocity and potential vorticity fields measured by altimetric imaging velocimetry in the rotating fluid, *Exp. Fluids*, **47**, 913.

Afanasyev, Y. D., and J. Wells (2005), Quasi-two-dimensional turbulence on the polar beta-plane: laboratory experiments, *Geoph. Astroph. Fluid Dyn.*, **99**, 1.

Berloff, P., I. Kamenkovich, and J. Pedlosky (2009a), A model of multiple zonal jets in the oceans: dynamical and kinematical analysis. *J. Phys. Oceanogr.*, **39**, 2711–2734.

Berloff, P., I. Kamenkovich, and J. Pedlosky (2009b) A mechanism of formation of multiple zonal jets in the oceans. *J. Fluid Mech.*, **628**, 395–425.

Callies, J., R. Ferrari, J. M. Klymak, and J. Gula (2015) Seasonality in submesoscale turbulence, *Nature Communications 6*, 6862, doi:10.1038/ncomms7862.

Centurioni, L. R., J. C. Ohlmann, and P. P. Niiler (2008), Permanent meanders in the California Current System, *J. Phys. Oceanogr.* **38**, 1690.

Danilov, S. D., and D. Gurarie (2000), Quasi-two-dimensional turbulence, *Physics-Uspekhi* **43**(9), 863.

Huang, H.-P., B. Galperin, and S. Sukoriansky (2001), Anisotropic spectra in two-dimensional turbulence on the surface of a rotating sphere, *Phys. Fluids* **13**, 225.

Ivanov, L. M., C. A. Collins, and T. M. Margolina (2009), System of quasi-zonal jets off California revealed from satellite altimetry, *Geoph. Res. Lett.* **36**, L03609.

Maximenko, N. A., B. Bang, and H. Sasaki (2005), Observational evidence of alternating zonal jets in the world ocean, *Geoph. Res. Lett.* **32**, L12607.

Maximenko, N. A., O. V. Melnichenko, P. P. Niiler, and H. Sasaki (2008), Stationary mesoscale jet-like features in the ocean, *Geoph. Res. Lett.* **35**, L08603 (2008).

McWilliams, J.C. (2006), Fundamentals of Geophysical Fluid Dynamics, Cambridge University Press.

Panetta, L., (1993) Zonal jets in wide baroclinically unstable regions: Persistence and scale selection. *J. Atmos.Sci.*, **50**, 2073–2106.

Qiu, B., S. Chen, P. Klein, H. Sasaki, and Y. Sasai (2014), Seasonal mesoscale and submesoscale eddy variability along the North Pacific Subtropical Countercurrent. *J. Phys. Oceanogr.*, 44(12), 3079-3098, doi:10.1038/ncomms6636.

Radko, T., D. Peixoto de Carvalho, and J. Flanagan, (2014) Nonlinear equilibration of baroclinic instability: the growth rate balance model. *J. Phys. Oceanogr.*, **44**, 1919–1940.

Read, P. L., Y. H. Yamazaki, S. R. Lewis, P. D. Williams, R. Wordsworth, K. Miki-Yamazaki, J. Sommeria, H. Didelle, and A. M. Fincham (2007), Dynamics of convectively driven banded jets in the laboratory," *J. Atmos. Sci*. 64(11), 4031.

Rhines, P. B. (1975), Waves and turbulence on a beta-plane, *J. Fluid Mech.* 69(3), 417.

Sasaki, H., P. Klein, B. Qiu, and Y. Sasai (2014), Impact of oceanic-scale interactions on the seasonal modulation of ocean dynamics by the atmosphere. *Nature Communications*, 5, 1-8, doi:10.1038/ncomms6636.

Slavin, A. G., and Y.D. Afanasyev (2012), Multiple zonal jets on the polar beta plane, *Phys. Fluids*, 24, 016603.

Smith, K. S. (2007), The Geography of linear baroclinic instability in Earth's oceans, *J. Marine Res.*, **65**, 655-683.

Smith, C. A., K. G. Speer, and R. W. Griffiths (2014), Multiple zonal jets in a differentially heated rotating annulus, *J. Phys. Oceanogr*., 44, 2273–2291.

Sokolov, S., and S. R. Rintoul (2007), Multiple jets of the Antarctic Circumpolar Current south of Australia. *J. Phys. Oceanogr*., 37, 1394–1412.

Thompson, A. F. (2008), The atmospheric ocean: eddies and jets in the Antarctic Circumpolar Current, *Phil. Trans. R. Soc. A* 366, 4529–4541.

Vallis, G. K., and M. E. Maltrud (1993), Generation of mean flows and jets on a beta-plane and over topography, J. *Phys. Oceanog*. **23**, 1346.

Williams, P. D. and Kelsall, C. W. (2015), The dynamics of baroclinic zonal jets, *J. Atm. Sci.*, 72(3), pp 1137-1151.

Zhang Y., and Y. D. Afanasyev (2014), Beta-plane turbulence: experiments with altimetry. *Phys. Fluids* **26**, 026602.

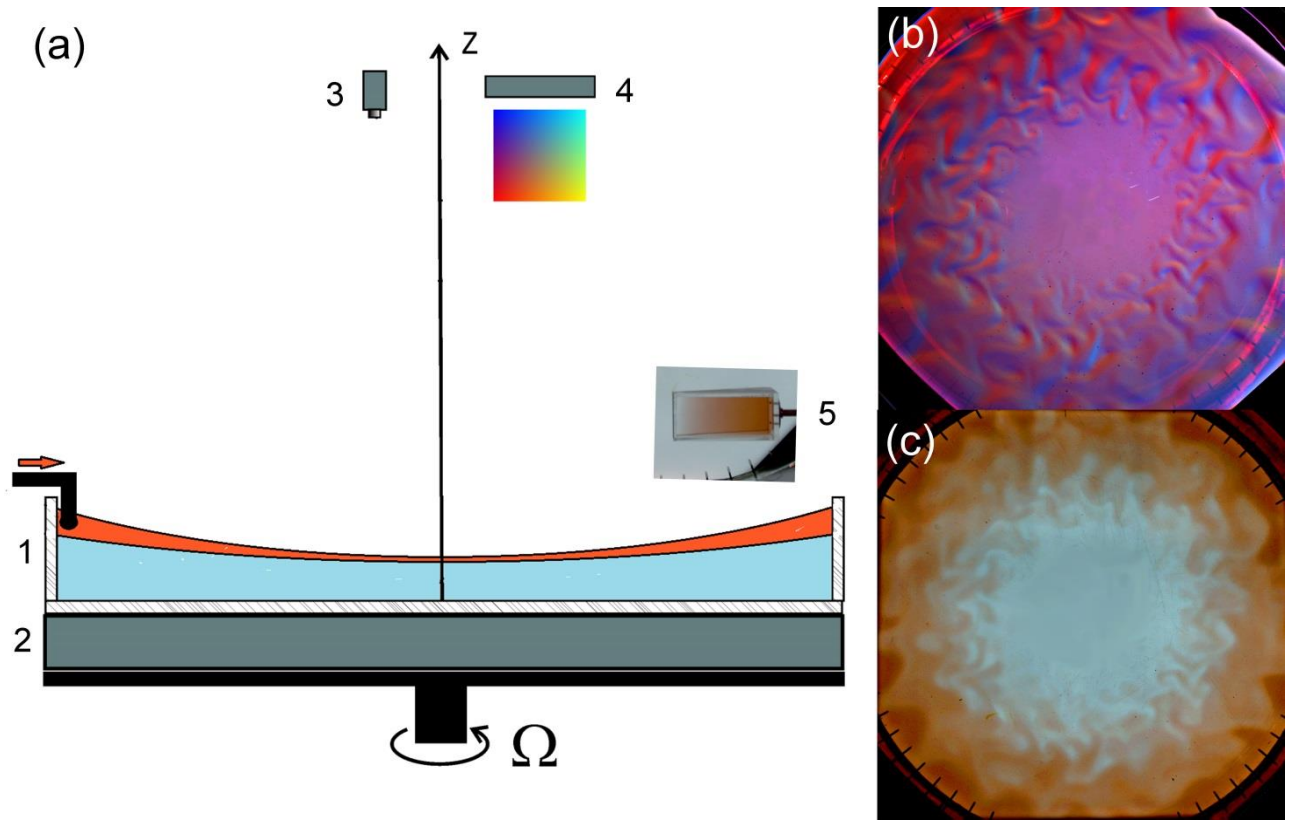

Figure 1. (a) Sketch of the experimental setup: (1) rotating tank, (2) light box for the optical thickness measurements, (3) video camera, (4) light source with color mask and (5) cuvette used for the calibration of the optical thickness method. Altimetric (b) and optical thickness (c) images.

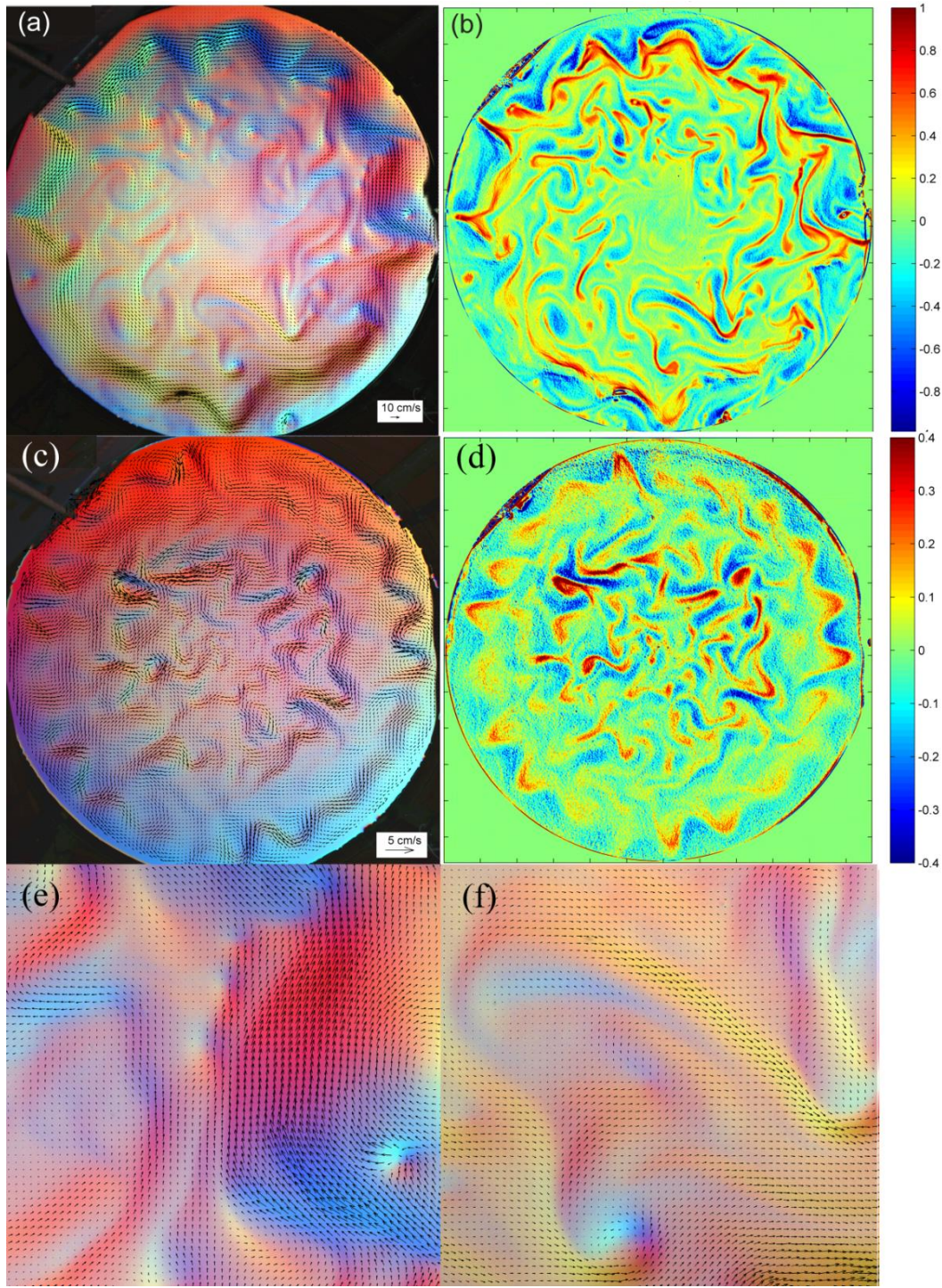

Figure 2. Baroclinically unstable flow visualized by the altimetry (AIV): (a) barotropic velocity (arrows) superposed on the altimetric image at $t = 115$ s; (b) relative vorticity of the barotropic component of the flow normalized by the Coriolis parameter; (c), (d) same as in (a) and (b) but at $t = 700$ s; (e) magnified view of the area of the flow in the top right part of the tank showing a chain of eddies within a filament; (f) flow in the bottom right part of the tank showing thin filaments protruding in the radial direction. Salinity difference between the layers is $S = 30$ ppt. The center of the tank corresponds to the North pole.
29

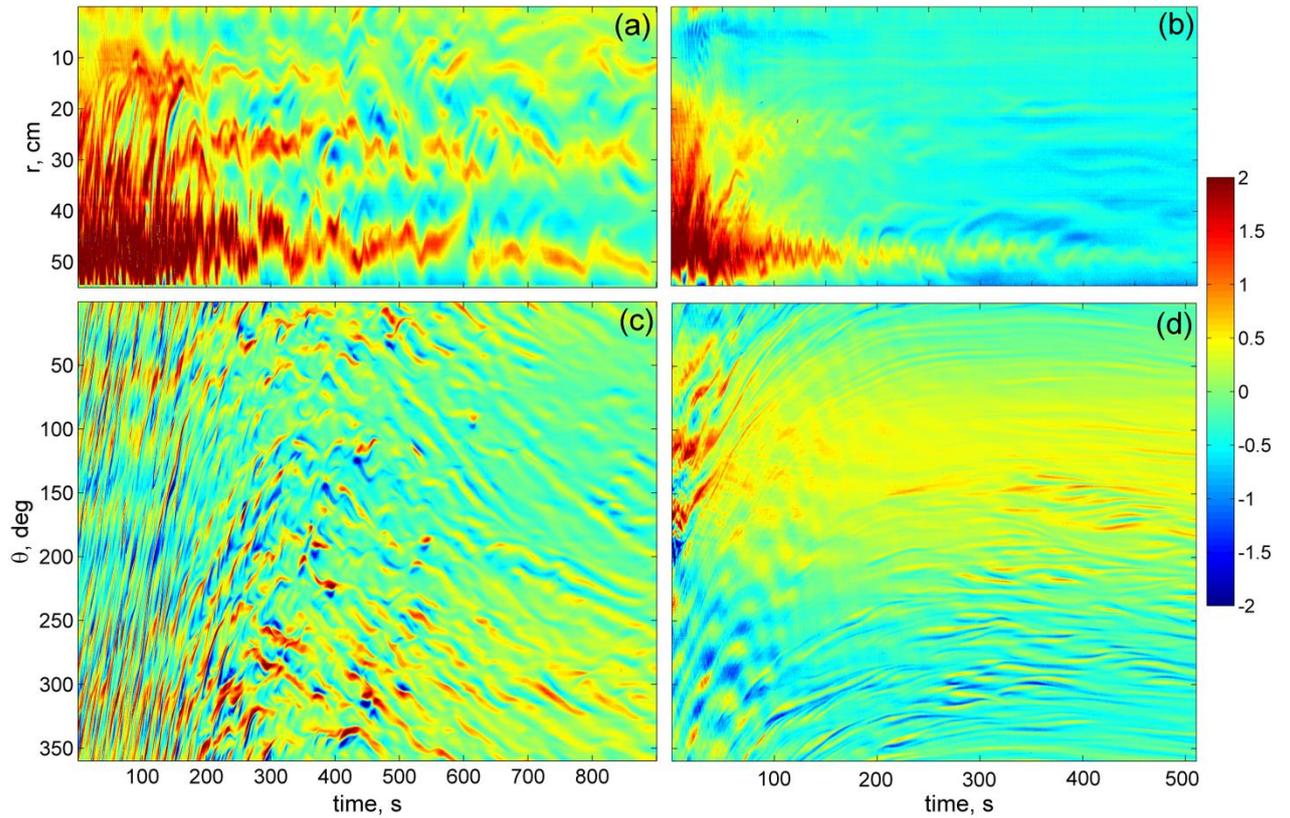

Figure 3. Hovmoeller (space-time) diagrams of azimuthal and radial velocities in two experiments: (a) and (b) azimuthal velocity measured along the radial direction in experiments with $S = 30$ ppt and $S = 10$ ppt respectively; (c) and (d) radial velocity measured along the azimuthal direction at $r = 2R/3$ in the same experiments. Color scale shows velocity in cm/s. Time starts after the forcing is stopped.

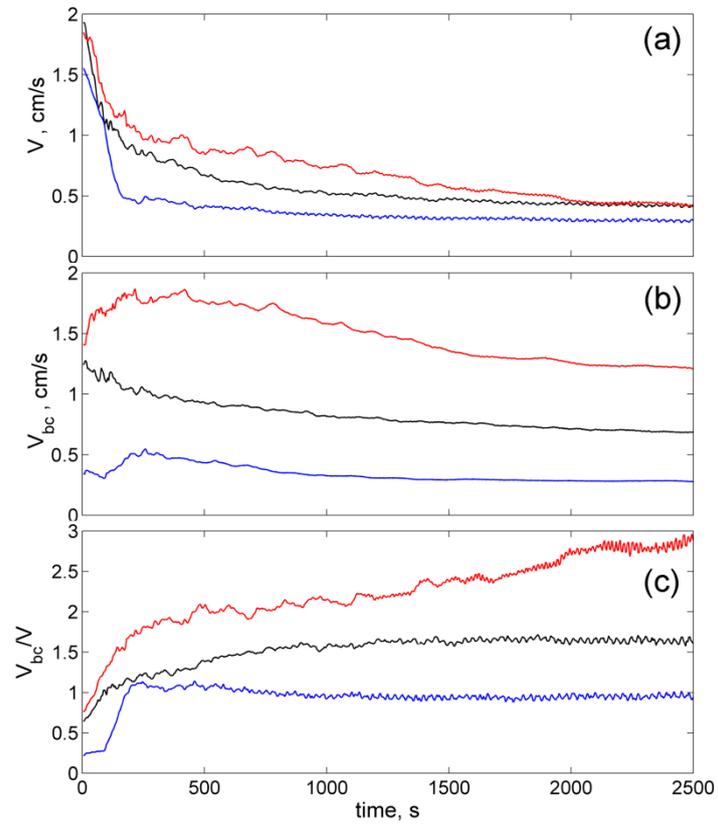

Figure 4. Root-mean-square barotropic (a) and baroclinic (b) velocities and their ratio (c) in three experiments with $S = 15$ (blue), 45 (black) and 65 ppt (red).

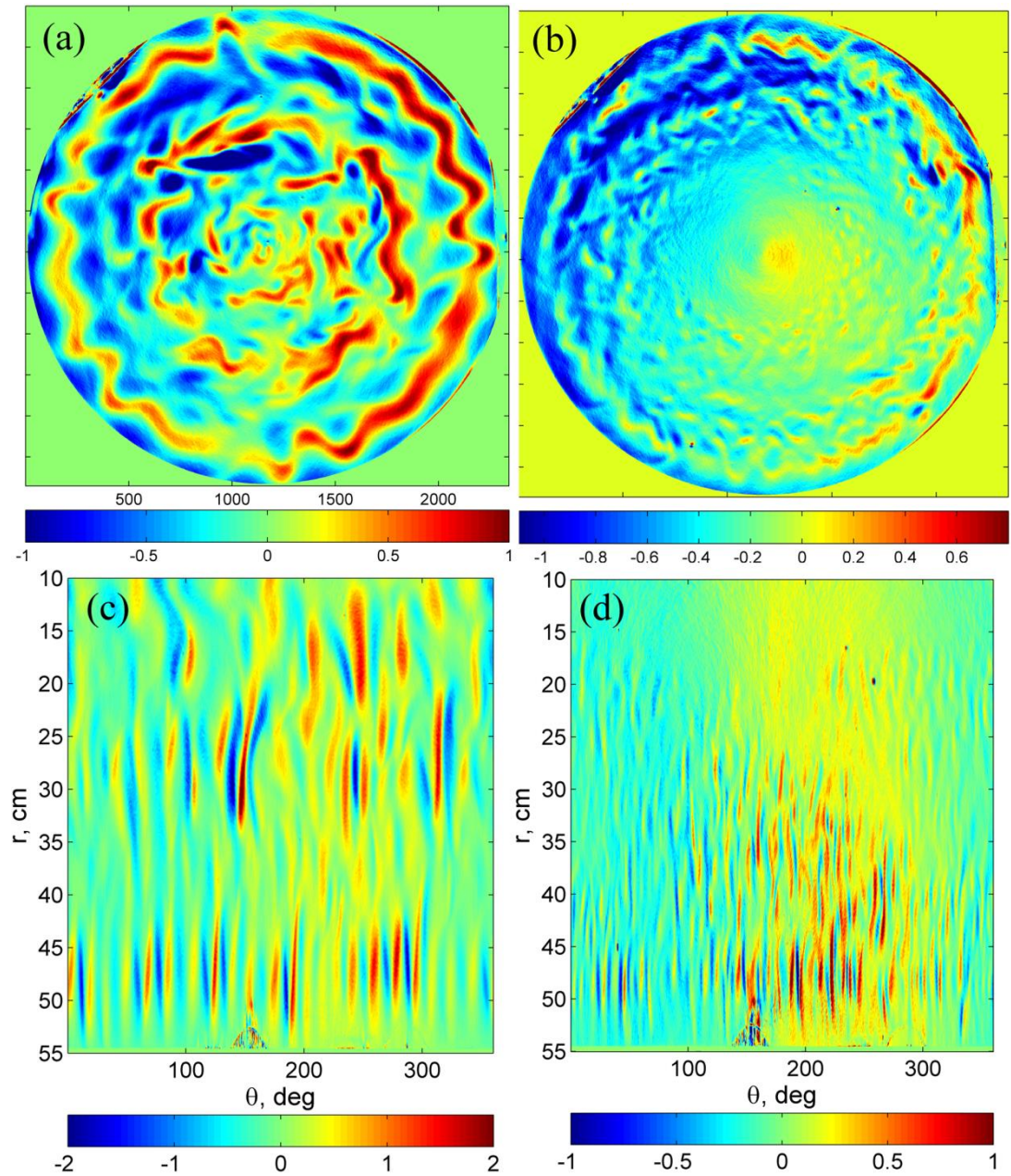

Figure 5. Azimuthal and radial velocity in two experiments: (a) and (b) azimuthal velocity in experiments with $S$ = 30 ppt and $S$ = 10 ppt respectively; (c) and (d) radial velocity measured in the same experiments in polar coordinate system. Color scales show velocity in cm/s.

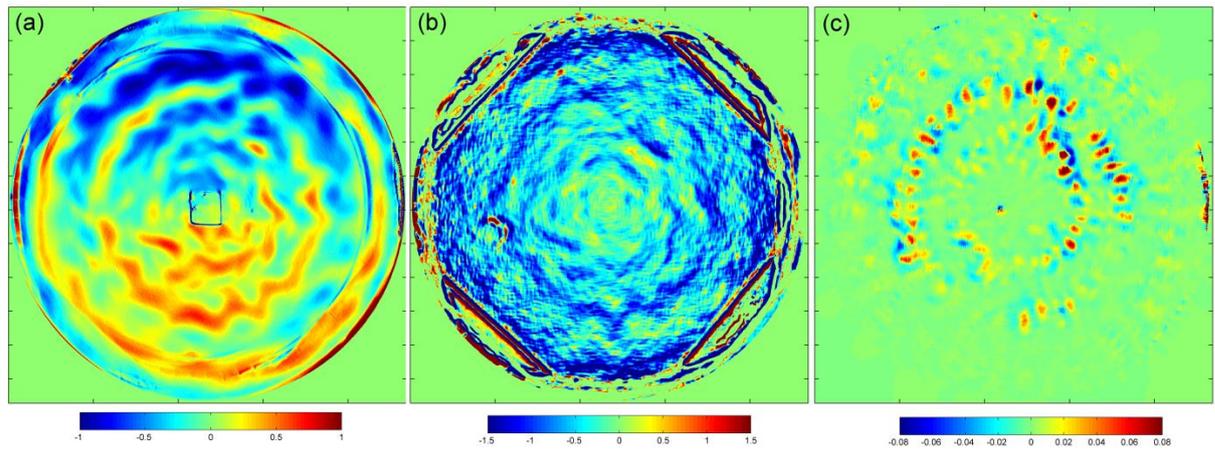

Figure 6. Azimuthal velocity and eddy forcing fields in the experiment with $S = 45$ ppt: (a) barotropic azimuthal velocity measured by AIV (upper layer velocity); (b) baroclinic azimuthal velocity measured by OTV; (c) azimuthal component of the Reynolds stress forcing in the upper layer. Color scales show the velocity in cm/s and the forcing in cm/s$^2$.

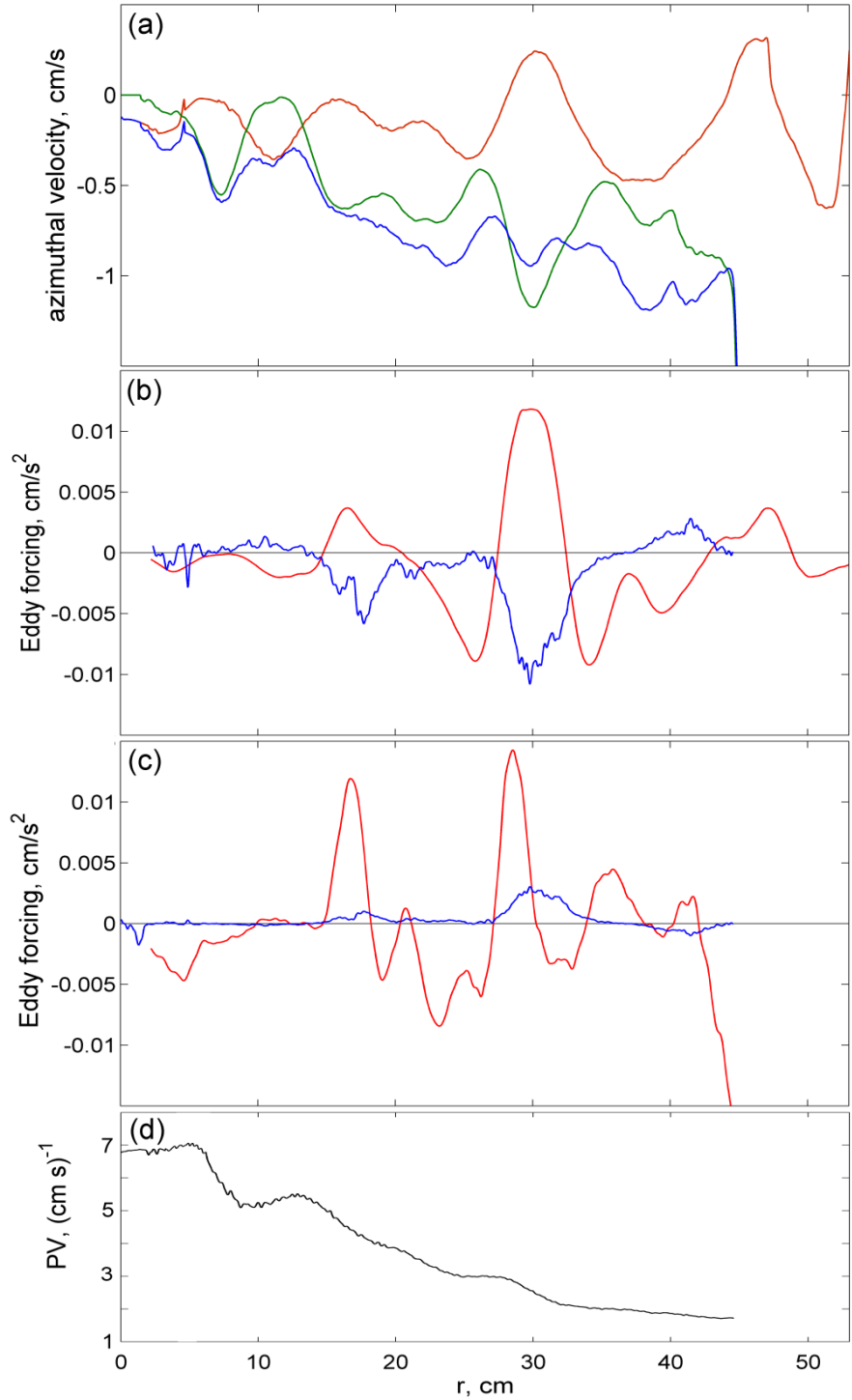

Figure 7. Profiles of the azimuthal velocity, eddy forcing and PV in the same experiment as in Figure 5: (a) radial profiles of the azimuthal barotropic/upper layer velocity $u_1 = U$ (red), azimuthal baroclinic velocity $U_{bc}$ (green) and azimuthal lower layer velocity $u_2 = U + U_{bc}$ (blue); (b) azimuthal component of the Reynolds stress

(red) and the form drag (blue) forcing in the upper layer; (c) same as in (b) but in the lower layer; (d) radial profile of PV in the upper layer.

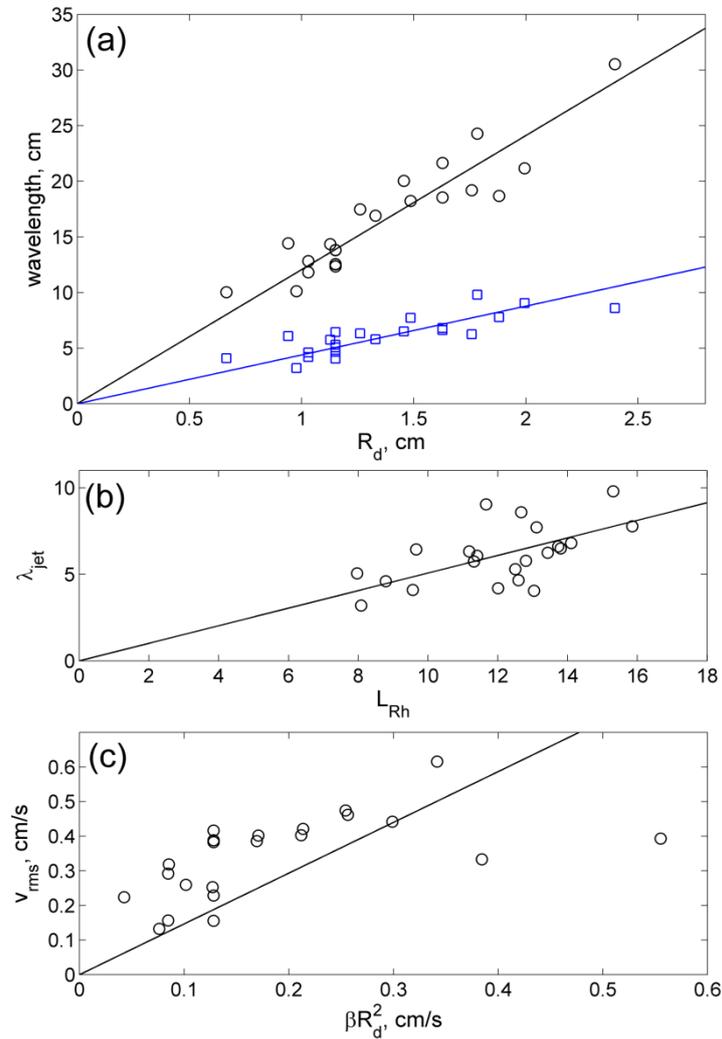

Figure 8. Characteristics of the flows measured in all experiments: (a) zonal wavelength, $\lambda_{zon}$, of the baroclinic meanders (blue squares) and meridional wavelength, $\lambda_{jet}$, of zonal jets (black circles) as functions of $R_d$; (b) $\lambda_{jet}$ (black circles) as a function of the modified Rhines scale, $L_{Rh}$; (c) rms radial velocity, $v_{rms}$, as a function of $\beta R_d^2$. Solid lines show a linear least squares fit of the data.

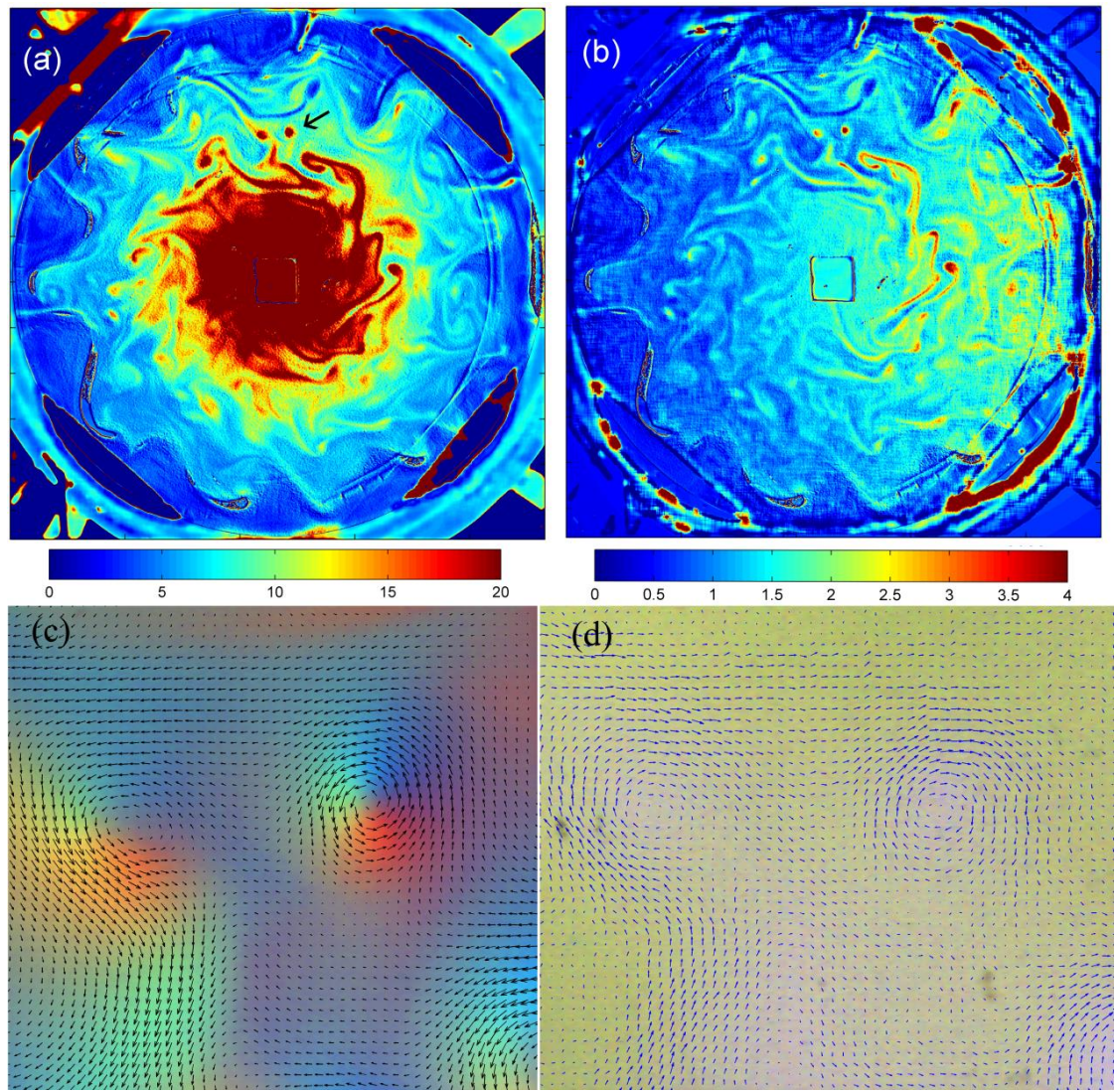

Figure 9. Flow fields in the experiment with $S = 45$ ppt soon after the forcing stopped: (a) potential vorticity in the upper layer, $q_1$; (b) potential vorticity in the lower layer, $q_2$; (c) barotropic velocity field superposed on the altimetric image in the eddy indicated by arrow in panel (a); (d) baroclinic velocity field superposed on the thickness image . Color scales show potential vorticity normalized by $f_0 H_0$.

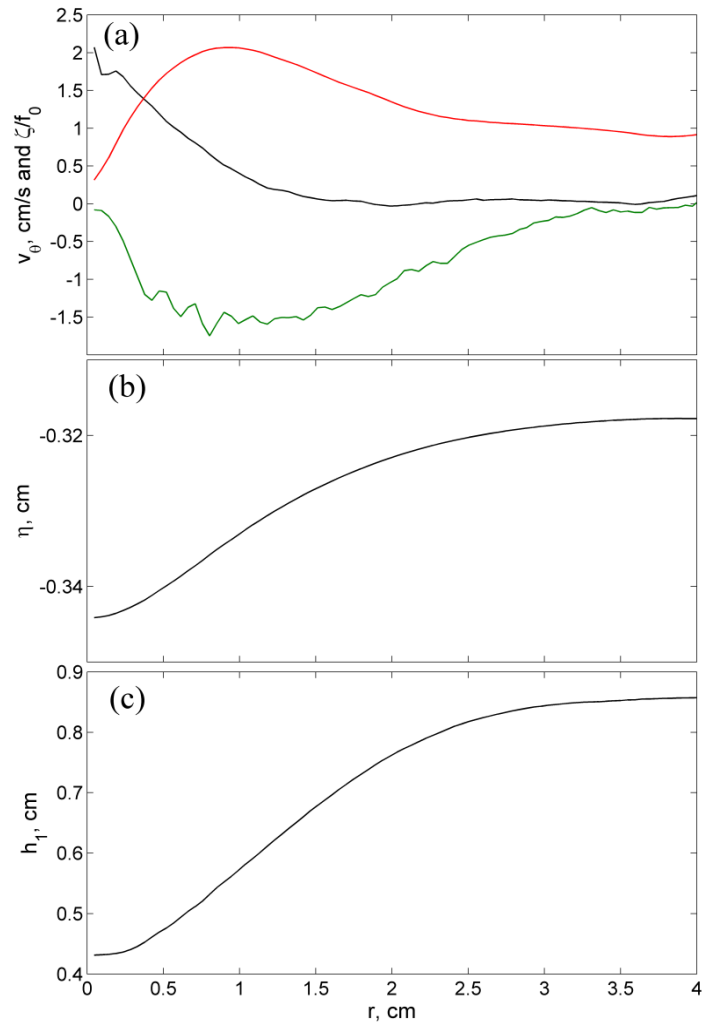

Figure 10. Radial profiles of the eddy characteristics: (a) azimuthal barotropic velocity (red), baroclinic velocity (green) and barotropic relative vorticity (black); (b) surface elevation; (c) upper layer thickness.